\documentclass{article}
\usepackage{amssymb}
\usepackage{amsmath}

\begin{document}

\begin{center}
Physics of Non-Inertial Reference Frames

Timur F. Kamalov

Physics Department, Moscow State Open University

Korchagina, 22, Moscow, 107996, Russia

E-mail: TimKamalov@mail.ru, TimKamalov@gmail.com
\end{center}

\textit{Physics of non-inertial reference frames is a generalizing of
Newton's laws to any reference frames. The first, Law of Kinematic in
non-inertial reference frames reads: the kinematic state of a body free of
forces conserves and determinates a constant n-th order derivative with
respect to time being equal in absolute value to an invariant of the
observer's reference frame. The second, Law of Dynamic extended Newton's
second law to non-inertial reference frames and also contains additional
variables there are higher derivatives of coordinates. Dynamics Law in
non-inertial reference frames reads: a force induces a change in the
kinematic state of the body and is proportional to the rate of its change.
It is mean that if the kinematic invariant of the reference frame is n-th
derivative with respect the time, then the dynamics of a body being affected
by the force F is described by the (n+1)-th differential equation. The
third, Law of Static in non-inertial reference frames reads: the sum of all
forces acting a body at rest is equal to zero.}



Newton's laws are valid in inertial reference frames, with the Lagrangian
being dependent on coordinates and their first derivatives (i.e.
velocities). A mathematician would call three Newton's laws an axiomatic of
physical theory. However there exists Ostrogradski's Canonical Formalism.
This mathematical description, in which we shall try to find a physical
meaning, will be called Ostrogradski's physics. For this we formulate two
postulates. The Lagrangian depends not only on coordinates and their first
derivatives (i.e. velocities) but also on higher derivatives of coordinates,
with he higher derivatives being considered independent variables. Really,
if the first derivative of coordinates can be independent of coordinates,
why cannot the higher derivatives? Repeating the well known procedure of
obtaining Euler-Lagrange's equation for such a Lagrangian, we shall obtain
Euler-Lagrange's equation with additional variables depending on the higher
derivatives of coordinates. Ostrogradski's physics is a physics of
non-inertial reference frames. This is the theory with a different
axiomatic, and hence it gives different results. What axiomatic of
non-inertial reference frames physics should be? The first axiom reads: free
particles conserves its kinematical states in any reference frame including
an inertial one. The second axiom generalized Newton's second law to any
reference frames and also contains additional variables there are higher
derivatives of coordinates and inertial forces, since Ostrogradski's physics
considers any reference frames including an inertial one. Newton's physics
is a particular case of Ostrogradski's Canonical Formalism for inertial
reference frames.

Classical physics usually considers the motion of bodies in inertial
reference frames. This is a simplified and approximate description of the
real pattern of the motion, as it is practically impossible to get an ideal
inertial reference frame. Actually in the any particular reference frame
there always exist minor influences due to any random fields. A simplified
consideration of the actual reference frame as an inertial one enables
derivation of motion equations, which are usually solved by means of the
traditional methods of mathematical physics. Then the uncertainty principle
is induced by the non-inertial character of the reference frame and
constitutes the expression of the error of measurement of coordinate and
momentum of the object under consideration and is a consequence of the
idealization of the problem being considered to an inertial reference frame.
In this case, one can assess the effect of inertial force in a non-inertial
reference frame through the Plank constant.

Let us consider the precise description of the dynamics of the motion of
bodies taking into account complex non-inertial nature of reference frames.
For this end, let us consider a body in a any reference frame, denoting the
actual position of the body as $r$, actual momentum as $p$ and time as $t$.
Then, expanding into Taylor series the function $r=r(t)$ and $p=p(t)$, we get

\begin{center}
\begin{equation}
r=r_{0}+vt+\frac{at^{2}}{2}+\frac{1}{3!}\overset{\cdot }{a}t^{3}+\frac{1}{4!}%
\overset{\cdot \cdot }{a}t^{4}+...+\frac{1}{n!}\overset{\cdot }{r}%
^{(n)}t^{n}+...  \label{eq:1}
\end{equation}

\begin{equation}
p=p_{0}+\overset{\cdot }{p}t+\frac{\overset{\cdot \cdot }{p}t^{2}}{2}+\frac{1%
}{3!}\overset{\cdot \cdot \cdot }{p}t^{3}+...+\frac{1}{n!}\overset{\cdot }{p}%
^{(n)}t^{n}+...  \label{eq. 1a}
\end{equation}
\end{center}

The kinematical state of the particle is define when any derivatives of
coordinates on time equal to zero, i.e. $\overset{\cdot }{r}^{(n)}=0$. Then
the derivative of coordinates on time $\overset{\cdot }{r}^{(n-1)}=const$ is
the kinematical invariant of the reference frame.

Let us compare this expansion with the well-known kinematical equation for
inertial reference frames of Newtonian physics relating the distance to the
acceleration $a$,%
\begin{equation}
r_{inertial}=r_{0}+vt+\frac{at^{2}}{2}  \label{eq:2}
\end{equation}

and the momentum

\begin{equation}
p_{inertial}=p_{0}.  \label{eq. 2a}
\end{equation}

Denoting the hidden (or correction and addition) variables accounting for
additional terms in any reference frames with respect to inertial ones as $%
\Delta r$ and $\Delta p$, we get%
\begin{equation}
\Delta r=\frac{1}{3!}\overset{\cdot }{a}t^{3}+\frac{1}{4!}\overset{\cdot
\cdot }{a}t^{4}+...+\frac{1}{n!}\overset{\cdot }{r}^{(n)}t^{n}+...
\label{eq:3}
\end{equation}

\begin{equation}
\Delta p=\overset{\cdot }{p}t+\frac{\overset{\cdot \cdot }{p}t^{2}}{2}+\frac{%
1}{3!}\overset{\cdot }{a}t^{3}+\frac{1}{4!}\overset{\cdot \cdot }{a}%
t^{4}+...+\frac{1}{n!}\overset{\cdot }{r}^{(n)}t^{n}+...  \label{eq. 3a}
\end{equation}

Then%
\begin{equation}
r_{any}=r_{inertial}+\Delta r  \label{eq. 4}
\end{equation}

\begin{equation}
p_{any}=p_{inertial}+\Delta p.  \label{eq. 4a}
\end{equation}

In this case, the measurement error of an experiment follows from
incompleteness of the description of sample particles in inertial reference
frames, as we assume the actual space-and-time to be a non-inertial
reference frame:%
\begin{equation}
(p_{any}-p_{inertial})(r_{any}-r_{inertial})\leq h,  \label{eq. 5}
\end{equation}

\begin{equation}
\Delta p\Delta r\leq h,  \label{eq. 5a}
\end{equation}

$h$ being the bound of correction's variables at the transformation of
inertial to any reference frames. Then comparing this in equation with the
uncertainty relation for $\Delta r=r-<r>$ and $\Delta p=p-<p>$ means
uncertainty of coordinates and momentums of the particle in the process of
measurements

\begin{equation}
\Delta p\Delta r\geq h,  \label{eq. 6}
\end{equation}

we can expect that $h$ is the some constant.

For non-inertial reference frames, the $h$ constant accounts for the effect
of the non-inertial space-and-time. Higher time derivatives of spatial
coordinates act as hidden variables complementing the description of sample
particles for inertial reference frames.

Newton's laws are valid in inertial reference frames with the Lagrangian $L$
is the function of only the coordinates and their first derivatives (i.e.
velocities), $L=L(t,r,\overset{\cdot }{r})$. For non-Newtonian physics in
non-inertial reference frames, the Lagrangian depends on the coordinates and
their higher derivatives as well as of the first one, i.e. $L=L(t,r,\overset{%
\cdot }{r},\overset{\cdot \cdot }{r},\overset{\cdot \cdot \cdot }{r},...,%
\overset{\cdot (n)}{r})$. Here the coordinates and their higher derivatives
is independent. Applying the principle of least action, we get [3]%
\begin{equation}
\delta S=\delta \int L(r,\overset{\cdot }{r},\overset{\cdot \cdot }{r},%
\overset{\cdot \cdot \cdot }{r},...,\overset{\cdot (n)}{r})dt=\int
\sum_{n=0}^{N}(-1)^{n}\frac{d^{n}}{dt^{n}}(\frac{\partial L}{\partial
\overset{\cdot (n)}{r}})\delta rdt=0\text{.}  \label{eq. 7}
\end{equation}%
Then, the Euler -- Lagrange function for complex non-inertial reference
frames takes on the form

\begin{center}
\begin{equation}
\frac{\partial L}{dr}-\frac{d}{dt}(\frac{\partial L}{\partial \overset{\cdot
}{r}})+\frac{d^{2}}{dt^{2}}(\frac{\partial L}{\partial \overset{\cdot \cdot }%
{r}})-\frac{d^{3}}{dt^{3}}(\frac{\partial L}{\partial \overset{\cdot \cdot
\cdot }{r}})+\frac{d^{4}}{dt^{4}}(\frac{\partial L}{\partial \overset{\cdot
(4)}{r}})+...=0  \label{eq. 8}
\end{equation}
\end{center}

It is the equation of the motion of particle with fee of forces influence in
non-inertial reference frames

\begin{center}
\begin{equation}
\sum_{n=0}^{N}(-1)^{n}\frac{d^{n}}{dt^{n}}(\frac{\partial L}{\partial
\overset{\cdot (n)}{r}})=0  \label{eq. 9}
\end{equation}
\end{center}

Let us consider in more detail this precise description of the dynamics of
body motion, taking into account of real reference systems. To describe the
extended dynamics of a body in an arbitrary coordinate system (corresponding
to any reference system) let us introduce concepts of kinematic state and
kinematic invariant of an arbitrary reference system.

\textbf{Definition}: \textit{the kinematic state of a body free of forces
determinates a constant }$n$\textit{-th order derivative with respect to
time being equal in absolute value to an invariant of the observer's
reference frame.}

Considering the dynamics of particles in any reference systems, we suggest
the following important laws.

\textbf{Law of Kinematic in non-inertial reference frames.} \textit{%
Kinematics Law in non-inertial reference frames reads: the kinematic state
of a body free of forces conserves and determinates a constant n-th order
derivative with respect to time being equal in absolute value to an
invariant of the observer's reference frame.} \textit{The kinematics state
of a body is defined if the n-th derivative of its coordinate with respect
to time is finite and equal to a negative value of the reference frame
invariant.} That is,

\begin{eqnarray}
q &=&q_{0}+\dot{q}t+\frac{1}{2!}\ddot{q}t^{2}+...+\frac{1}{n!}\dot{q}%
^{(n)}t^{n},  \label{eq. 10} \\
\frac{d^{n}q}{dt^{n}} &=&\dot{q}^{(n)}=const.
\end{eqnarray}

The acceleration for a body which free from forces influence is a constant
for the observer in the uniformed reference frame. In this case the
acceleration is defining the kinematic state of the body. The Law of
Kinematics is the kind of equivalence principle and expresses the extension
of the First Newton's Law to non-inertial reference frames. We can find so
many planets in the Universe for this case. In the extended model of
dynamics, the transition from a reference frame to another one is defined
the transformation of reference frames as

\begin{eqnarray}
q^{\prime } &=&q_{0}+\dot{q}t+\frac{1}{2!}\ddot{q}t^{2}+...+\frac{1}{n!}\dot{%
q}^{(n)}t^{n}  \label{eq. 11} \\
t^{\prime } &=&t.
\end{eqnarray}

In this case Taylor's series decomposition of the coordinate must be
convergence.

\textbf{Law of Dynamic in non-inertial reference frames.} \textit{The force
acting to the particle equal to the velocity of changing the kinematics
state of the particle. If the kinematic invariant of an arbitrary reference
frame is }$n$\textit{-th time derivative of body coordinate, then the body
dynamics with influence of the force }$F(t,q,\dot{q},\ddot{q},...,\dot{q}%
^{(n)})$\textit{\ is described with the differential equation of the order (}%
$n+1)$\textit{:}

\begin{equation}
\alpha _{n+1}\dot{q}^{(n+1)}+...+\alpha _{2}\ddot{q}+\alpha _{1}\dot{q}%
+\alpha _{0}q=F(t,q,\dot{q},\ddot{q},...,\dot{q}^{(n)}).  \label{eq. 12}
\end{equation}

\textit{Here }$\alpha _{n}$\textit{- some constants.}

Here (12) is the modification of the Newton's Second Law [1] for the general
case of non-inertial reference frames. Odd derivatives correspond to losses
(friction or radiation) and describe irreversible cases for open systems not
satisfying variational principles of mechanics.

\textbf{Law of Static in non-inertial reference frames. }\textit{In
arbitrary reference frames the sum of forces which action to the statics
particle is equal to zero.}

The Generalized Poisson's equation for the scalar potential $\varphi $ of
gravitational field in this case from the sources with density distribution
of the source $\rho $ and factor $\varkappa $ depending on the system of
units shall take on the form

\begin{center}
\begin{equation}
\sum_{n=0}^{N}{}^{n}\frac{\partial \varphi }{\partial \overset{\cdot (n)}{r}}%
=\varkappa \rho  \label{eq. 13}
\end{equation}
\end{center}

or, in our case, Generalized Poisson's equation is

\begin{center}
$\sum_{n=0}^{N}\overset{\cdot (n)}{\nabla }\varphi =\varkappa \rho $.
\end{center}

Than the solution of Generalized Poisson's equation is

\begin{center}
\begin{equation}
\varphi (t,r(t))=\sum_{n=0}^{N}\varphi _{0}\exp (-k/\overset{\cdot (n)}{r}%
(t)).  \label{eq, 14}
\end{equation}
\end{center}

In the particular, for discussion the speculation of gravity the
gravitational field the potential for example is

\begin{center}
\begin{equation}
\varphi =\varphi _{0}\exp (-k/r)=\frac{GM}{r}\exp (-k/r),  \label{eq. 15}
\end{equation}
\end{center}

where $\varphi $- potential, $G$- gravitational constant, $k$ - unknown
constant or the scale of the interaction, $M=\int \rho dv$ - mass, $%
r=x-x_{0}<<1$, $x$ and $x_{0}$- coordinates. The constant $k$ is unknown,
but if $k$ is equal to the Plank constant $l_{p}=10^{-33}$ cm than this
potential is always the same as Newtonian potential $\varphi =GM/r$. If
constant $k$ is equal to the size of nuclear $k=10^{-15}m$ than the
gravitational force is equal to nuclei forces because at the small distant
gravitational forces is change on exponential law and be strong.

From this paper follow, that the phase space of coordinates and there
multiple derivative gives the modification of the Newton's formula for the
small scales for gravitational potential $\varphi $ of two mass $m$\ is

\begin{center}
$\varphi =\frac{Gm}{r}(1-b\frac{k}{r}+c\frac{k^{2}}{r^{2}}-d\frac{k^{3}}{%
r^{3}}+...)$,
\end{center}

where $a$,$b$,$c$,... - constants. In the particular case when $r>>k$\ from
(14) gravitational potential $\varphi $ is

\begin{equation}
\varphi =\frac{GM}{r}(1-\frac{k}{r}+\frac{k^{2}}{r^{2}}-\frac{k^{3}}{r^{3}}%
+...)  \label{eq. 16}
\end{equation}

Here $k$ is the unknown constant which have the seance of distance. For
example, if $k\thicksim 10^{-15}m$ and $r>k$\ we have always Newtonian low.

For long distances $r>>k$, we have the equation for the Newtonian
gravitational potential $\varphi _{0}=Gm\frac{1}{r}$. For the distance $r<k$
the gravitational potential $\varphi $ is a strong and in this case we can
compare the gravitational force with the nuclear force. This modification of
the Newton's gravitation law we can consider on the case of the dark matter.

For particle described by the generalized Hamilton function at small
distances, i.e. when the series diverges, there shall be much stronger
forces acting than it is usually considered in calculations employing the
Hamilton function. This theory of short-range interaction explains
interaction of bodies at small distances and refines the description of
their interaction in case of their increase. It can be supposed that this
method can be applied to cases when the force of gravitational attraction of
particles described by the Hamilton function, at low distances.

Denoting the addition energy brought about by the non-inertial reference
frame as $Q$ and the constant coefficients as $\alpha _{i}$, we get for the
total energy $E$, potential energy $V$ and kinetic energy $W$ the following
expressions:

\begin{center}
$E=\alpha _{0}r^{2}+\alpha _{1}\overset{\cdot }{r}^{2}+\alpha _{2}\overset{%
\cdot \cdot }{r}^{2}+\alpha _{3}\overset{\cdot \cdot \cdot }{r}%
^{2}+...+\alpha _{n}\overset{\cdot (n)}{r}^{2}+...$

$E=V+W+Q$

$V=\alpha _{0}r^{2}$

$W=\alpha _{1}\overset{\cdot }{r}^{2}$

$Q=\alpha _{2}\overset{\cdot \cdot }{r}^{2}+\alpha _{3}\overset{\cdot \cdot
\cdot }{r}^{2}+...+\alpha _{n}\overset{\cdot n}{r}^{2}+...$
\end{center}

Generalized Jacobi-Hamilton equation in the weak \ non-inertial reference
frame for the action function takes on the form:

\begin{center}
\begin{equation}
-\frac{\partial S}{\partial t}=\frac{(\nabla S)^{2}}{2m}+V+Q,  \label{eq. 17}
\end{equation}
\end{center}

and let us call $Q$ the quantum potential. Here, is the velocity $v=\frac{%
\partial S}{\partial t}=\frac{\nabla S}{m}$, and the acceleration $a=\overset%
{\cdot }{v}=\frac{\nabla \overset{\cdot }{S}}{m}=\frac{\nabla ^{2}S}{m}$,
where is the continuity equation $\frac{\partial v}{\partial t}+\nabla v=0$
for the vector $v$. Here is $\overset{\cdot \cdot }{v}=\nabla \overset{\cdot
\cdot }{S}$.

In the first approximation $Q\approx \alpha _{2}\frac{\nabla ^{2}S}{m}=-%
\frac{i\hbar }{2m}\nabla ^{2}S.$ (the constant is chosen as $\alpha _{2}=%
\frac{i\hbar }{2}$) is Bohm's quantum potential. Hence, we get the Schr\"{e}%
dinger equation in the first approximation for the function $\psi =Ae^{\frac{%
i}{\hbar }S}$ from the equation (17) [4]

\begin{center}
$-\frac{\partial S}{\partial t}=\frac{(\nabla S)^{2}}{2m}+V-\frac{i\hbar }{2m%
}\nabla ^{2}S$
\end{center}

Here we complete the Classical Physics with the hidden variables of the real
non-inertial reference frames. In this case the weak influence of inertial
forces in non-inertial reference frame define the quantum behavior of
particles.

\begin{center}
\textit{References}
\end{center}

[1] Newton I., Philosophiae naturalis principia mathematica, 1687.

[2] Lagrange J.I., Mecanique analitique. Paris, De Saint, 1788.

[3] Ostrogradskii M., Met. De l'Acad. De St.-Peterburg, v. 6, p. 385, 1850.

[4] Bohm D., Phys. Rev. 85, 166, 1952; Phys. Rev. 85, 180, 1952; Phys. Rev.,
87, 389, 1952.

\end{document}